\documentclass[twocolumn,10pt]{article}

\usepackage{preprint}
\usepackage{amsmath,amsfonts,amsthm}
\usepackage{graphicx}
\usepackage{tikz}
\usepackage{url}
\usepackage{pgfplots}
\usepackage{textcomp}
\usepackage{xcolor}
\def\BibTeX{{\rm B\kern-.05em{\sc i\kern-.025em b}\kern-.08em
    T\kern-.1667em\lower.7ex\hbox{E}\kern-.125emX}}

\usepackage{xspace}
\usepackage{subfig}
\usepackage{algorithm}
\usepackage{algpseudocode}

\usepackage[inline]{enumitem} \setlist{noitemsep,nolistsep}
\usepackage{units}  

\usepackage{fnpct}

\usepackage[capitalize,noabbrev]{cleveref}

\newtheorem{theorem}{Theorem}
\newtheorem{lemma}[theorem]{Lemma}

\newtheorem*{remark*}{Remark}

\newcommand{\rindx}{r}

\newcommand{\col}[2]{#1[#2]}
\newcommand{\row}[2]{#1[#2,\cdot]}
\newcommand{\low}{\mathtt{{low}}}

\newcommand{\val}{f}
\newcommand{\uid}{\mathtt{uid}}
\newcommand{\RR}{\mathbb{R}}
\newcommand{\ZZ}{\mathbb{Z}}

\newcommand{\dint}[1]{\mathring{D}_{#1}}
\newcommand{\rint}[1]{\mathring{R}_{#1}}
\newcommand{\dsp}[1][i]{{D}^S_{#1}}
\newcommand{\dspi}[1][i]{{\mathring{D}}^{S}_{#1}}
\newcommand{\dsps}[1][i]{{D}^{SS}_{#1}}

\newcommand{\rspi}[1]{\mathring{R}^S_{#1}}
\newcommand{\rsps}[1]{{R}^{SS}_{#1}}
\newcommand{\rseg}[1]{[s_{#1-1}, s_{#1})}

\newcommand{\rfin}{R^F}

\newcommand{\piv}[1]{\mathtt{pivots[#1]}}
\newcommand{\npiv}{\mathtt{pivots}}

\newcommand{\updated}{\mathtt{updated}}

\algtext*{EndWhile}
\algtext*{EndIf}%
\algtext*{EndFor}%
\algrenewcommand\algorithmiccomment[1]{// {\itshape #1}}

\newcommand{\cadmus}{{\textsc{Cadmus}}\xspace}
\newcommand{\dipha}{{\textsc{DIPHA}}\xspace}
\newcommand{\diy}{{\textsc{DIY}}\xspace}

\newcommand{\Remark}[1]{}

\begin{document}
\pgfplotsset{compat=1.15}

\title{Distributed Computation of Persistent Cohomology}



\author{Arnur Nigmetov, Dmitriy Morozov \\ Lawrence Berkeley National Laboratory \\ anigmetov@lbl.gov, dmorozov@lbl.gov }


\twocolumn[
\maketitle

\begin{abstract}
    \noindent
    Persistent (co)homology is a central construction in topological data
    analysis, where it is used to quantify prominence of features in data to
    produce stable descriptors suitable for downstream analysis.
    Persistence is challenging to compute in parallel because it relies on
    global connectivity of the data.  We propose a new algorithm to compute
    persistent cohomology in the distributed setting.  It combines domain and
    range partitioning. The former is used to reduce and sparsify the coboundary
    matrix locally.  After this initial local reduction, we redistribute the
    matrix across processors for the global reduction.  We experimentally
    compare our cohomology algorithm with \dipha, the only publicly available
    code for distributed computation of persistent (co)homology; our algorithm
    demonstrates a significant improvement in strong scaling.
\end{abstract}
\vspace{1cm}
]

\section{Introduction}

\Remark{TDA and applications.}
Topological data analysis is a research area at the intersection of
computational geometry and algebraic topology. It aims to understand the ``shape
of data'' by identifying prominent topological features that are stable to
perturbations. One of its central constructions is persistent
(co)homology~\cite{ph-survey,ph-book-chapter},
which achieves this aim by examining the data across a range of scales and
keeping track of the values where features are born and die. By pairing these
events, it generates a \emph{persistence diagram}, which is a stable descriptor
that summarizes the distribution of topological features as a point set in the
plane.

Persistence has found numerous applications. In cosmology, it's been used to
describe the shape of the Cosmic Web~\cite{PE+17}---the distribution of matter forming the
large-scale structure of the Universe---and thus to compare different
cosmological models. In materials science, persistence has been used to detect
cavities and channels in nanoporous materials, information that was in turn used
as an input for a machine learning algorithm to predict adsorption of a
greenhouse gas~\cite{KHM20}. In machine learning, persistence has been used to regularize the
training loss~\cite{chao-yusu,pso} to reduce topological complexity of the decision boundary, and
thus minimize the model overfitting to outliers.

Some of the applications of persistence require processing very large data sets.
For example, cosmological simulations are some of the largest users of modern
supercomputers, modeling domains on the order of $8,192^3$ grid cells.  In other
applications, even when the input data set is small, the combinatorial
construction of simplicial complexes required to represent it for the
persistence computaiton explodes the input complexity, scaling exponentially
with homological dimension. These observations highlight the need for
distributed algorithms and software to compute persistence, both to improve the
running times and to gain access to the distributed (multi-node) memory.

Two types of distributed algorithms have been proposed in the literature.  The
first~\cite{lewis2015parallel} partitions data by domain, processes as much of
the computation locally as possible, and then performs a reduction among all
processes to deduce the global connectivty of the topological features. The
global reduction uses a gluing procedure called \emph{Mayer--Vietoris blowup
complex}, which adds extra cells to represent the structure of the intersecting
regions of the domain. Algebraically, it can be interpreted as iterating through
the Mayer--Vieotirs spectral sequence~\cite{Casas2019}.
The approach suffers from Amdahl's law: for small number of processors, the
computation scales very well, since the local work dominates, but the running
time quickly plateaus as the global gluing dominates the work and additional
processes cannot help~\cite{lewis2015parallel}.

The second approach partitions the data by function value. The algorithm is
based on the spectral sequence of the filtration~\cite{ph-survey}; it is
implemented in the only publicly available code for this problem,
\dipha~\cite{bauer2014distributed,dipha-repo}.
The input matrix is reduced by blocks, moving away from the diagonal. This
guarantees that each process only needs to access the data in the range it is
responsible for and can pass the data to its neighbors, when the reduction
leaves its local range.

Other techniques are available in shared memory, for example, identification of
apparent pairs~\cite{MST17,hypha} or waitfree column reduction that uses atomic
operations for synchronization~\cite{morozov2020towards}, but these do not help in the distributed
memory setting.

Our contribution is three-fold.
\begin{enumerate}[nosep]
    \item
        We show that if one switches from persistent homology to persistent
        cohomology, the coboundary matrix that serves as input to the
        computation has exactly the same structure as the boundary matrix of the
        blowup complex. Thus, the algorithm of \cite{lewis2015parallel} can be
        applied to it directly, and we use its first round to locally reduce
        data partitioned by the domain.
    \item
        We combine the two approaches and after one round of local domain
        reduction, we switch to the spectral sequence algorithm to reduce the
        data globally.
    \item
        We demonstrate that the combined approach scales significantly better
        than the persistent cohomology implementation in \dipha.
\end{enumerate}

\section{Background}
\subsection{Simplicial Complexes}
We start with a finite \emph{vertex set} $V$.
For instance, $V$ can be the set of nodes of a grid, on which a numerical
simulation evaluates some function $f$.
A $k$-\textit{simplex} is a set of $k+1$ vertices.
If $(k-1)$-simplex $\tau$ is a subset of $k$-simplex $\sigma$, $\tau \subset
\sigma$, then $\tau$ is called a \textit{facet} of $\sigma$.
The set of all facets of $\sigma$ forms its \textit{boundary},
denoted $\partial \sigma$. For simplicity, we work with mod 2 coefficient (in
$\ZZ/2\ZZ$), and we treat the boundary as a formal sum:
\[
    \partial \sigma = \sum_{\tau \subset \sigma, |\sigma \setminus \tau| = 1} \tau.
\]
The set of simplices $K$ is called a \textit{simplicial complex}, if for each $\sigma \in K$
it also contains all facets of $\sigma$ (and hence, recursively, all subsets of $\sigma$).
A \textit{filtration} of $K$ is a function on the simplicial complex,
$\val \colon K \to \RR$, such that $\val(\tau) \leq \val(\sigma)$ whenever $\tau \subset \sigma$.
We call $\val(\sigma)$ the \textit{filtration value} of $\sigma$. 

A specific construction, common for scientific computing data, that we use in our experiments is called
a \textit{lower-star} filtration. It extends a function $\hat{\val}: V \to \RR$ defined
on the vertices (e.g., of a simulation domain) to a filtration $\val: K \to \RR$,
defined on all the simplices (in the triangulation of the domain).
It assigns to each simplex the maximum value of its vertices,
\[
    \val(\sigma) = \max(\hat\val(v_0), \hat\val(v_1), \dots, \hat\val(v_k)).
\]
Here $\sigma = \{ v_0, \dots, v_k \} $.
Lower-star filtrations are used to
to capture the topology of sublevel sets, $\val^{-1}(-\infty,a]$ of scalar function.

Lower-star filtrations highlight that $\val$ need not be injective. However,
the computation of persistence requires a total order.
So we assume that all values of $\val$ are distinct, and every simplex is
uniquely determined by its value $\val(\sigma)$. To achieve this in practice, we
assign a unique identifier $\uid$ to each simplex and break ties by comparing
the pairs $(\val(\sigma), \uid(\sigma))$ lexicographically.

\subsection{Coboundary matrix}
For a fixed simplicial complex $K$, simplex $\tau$ is in the coboundary of
$\sigma$ if and only if $\sigma$ is in the boundary of $\tau$.
The set of all simplices $\tau \in K$ such that
$\sigma \in \partial \tau$ is called the \textit{coboundary} of $\sigma$.
Suppose that we have a total order on the simplices of $K$ (given by some
filtration). We enumerate
the simplices of $K$ in \textbf{reverse} filtration order and define the \textit{coboundary matrix} of $K$ by the formula
$D[i, j] = 1$ if and only if the $j$-th simplex is in the coboundary of the $i$-th simplex (in other words,
the \textit{rows} of $D$ encode the boundaries).
It is often convenient to extend the notation and use simplices themselves to
refer to the matrix entries:
$D[\sigma, \tau]$ means $D[i, j]$ when $\sigma$ has index $i$
and $\tau$ has index $j$ in the reverse filtration order.

\begin{remark*}
    We work with the most common case of $\ZZ/2\ZZ$ coefficients, which allows us
    to treat rows and column of all our matrices as sets of simplices. The theory of persistent
    homology is valid for arbitrary fields, and all algorithms in the paper
    can be trivially re-written for any $\ZZ/p\ZZ$ ($p$ prime).
\end{remark*}

\subsection{Sequential algorithm}
We write $\col{A}{k}$ to denote the $k$-th column of matrix $A$
and $\row{A}{k}$ to denote the $k$-th row of $A$.
The index of the last non-zero entry of a column is denoted $\low(\col{A}{k})$; for zero columns,
$\low$ is undefined. We say that two columns have a \textit{collision}, if they have the same $\low$.
Matrix $A$ is called \textit{reduced}, if there are no collisions. A \textit{valid} column
operation is adding a column from left to right: $\col{A}{k} \gets \col{A}{k} +
\col{A}{k'}$ when $k' < k$.
We usually perform this operation, if columns $k$ and $k'$ have a collision.
In this case,
this operation eliminates the collision, and either $\col{A}{k} = 0$ or $\low(\col{A}{k}) < \low(\col{A}{k'})$.
A \textit{reduction} is any sequence of valid operations that brings matrix $A$ to a reduced form.

\cref{alg:reduction_algorithm} presents a pseudocode for a generic reduction.
As written, it is not sufficiently detailed to be an algorithm:
we do not specify the order in which
\textbf{for all} iterates over the columns $\col{R}{i}$.
Nor do we specify a concrete choice of a column to the left $\col{R}{j}$
that has a collision with $\col{R}{i}$ (there can be more than one).

The usual way to check existence of column $\col{R}{j}$ that
has a collision with $\col{R}{i}$ (current column being reduced)
is to maintain a table of \textit{pivots}: given
a row index $a$, the entry $\piv{a}$
contains the index of a column with $\low(\col{R}{\piv{a}}) = a$
If no such column has been seen by the algorithm, $\piv{a}=-1$.
If we process all columns from left to right and,
once a column cannot be further reduced, mark it as a pivot,
we obtain the standard reduction algorithm \cite{edelsbrunner2002topological}.
Its pseudocode is in \cref{alg:elz}.

From a reduced matrix, one obtains a \textit{persistence pairing}: simplices
$\sigma$ and $\tau$ form a pair, if $\low(\col{R}{\sigma}) = \tau$.
Simplex $\sigma$ is called \textit{negative} and simplex $\tau$ is \textit{positive}.
Every simplex can appear in at most one pair; in particular, if $\tau$ is positive,
then its column $\col{R}{\tau}$ must be $0$\footnote{This follows from the fundamental property of the (co)boundary operator: its composition with itself is $0$.}.
The \textit{persistence diagram} of $K$ in dimension $d$ is defined
as the multi-set of all pairs $(\val(\sigma), \val(\tau)) \in \RR^2$ such that
simplices $\sigma$ and $\tau$ are paired, $\sigma$ is a $d$-simplex (hence $\tau$ must
be a $d+1$-simplex) and $\val(\sigma) \neq \val(\tau)$.
Although a reduced matrix is not unique, the pivots are, and so all reduced
matrices produce the same persistence pairing~\cite{vineyards}.

\begin{algorithm}
\caption{Reduction algorithm.}
\label{alg:reduction_algorithm}
\begin{algorithmic}[1]
    \Function{ReduceMatrix}{$D$}
        \State{$R \gets D$}
        \ForAll{column $\col{R}{i}$ of $R$}
            \While{$\col{R}{i} \neq 0$}
                \While{ $\exists j < i: \low(\col{R}{j}) = \low(\col{R}{i})$ }
                    \State {$\col{R}{i} \gets \col{R}{i} + \col{R}{j}$}
                \EndWhile
            \EndWhile
        \EndFor
        \State {\Return {$R$}}
    \EndFunction
\end{algorithmic}
\end{algorithm}

\begin{algorithm}
\caption{Reduction algorithm.}
\label{alg:elz}
\begin{algorithmic}[1]
    \Function{ReduceMatrixELZ}{$D$}
        \State{$R \gets D$}
        \State{$m, n \gets \mbox{number of rows and columns of $R$}$}
        \State{$\npiv = [-1, -1, \dots, -1]$} \Comment{array of length m}
        \For{$c \in [1, 2, \dots, n]$}
            \While{$\col{R}{c} \neq 0$}
                \State{$\ell \gets \low(\col{R}{c})$}
                \State{$p \gets \piv{\ell}$}
                \If {$p \neq -1$}
                    \State {$\col{R}{c} = \col{R}{c} + \col{R}{p}$}
                \Else
                    \State{$\piv{\ell} \gets c$}
                    \State{\textbf{break}}
                \EndIf
            \EndWhile
        \EndFor
        \State{\Return{$R$}}
    \EndFunction
\end{algorithmic}
\end{algorithm}

\subsection{Clearing}
Clearing was first suggested in \cite{chen2011persistent}.
The idea follows from the structure of the (co)boundary matrix:
if we know that a simplex $\sigma$ is positive,
then we do not need to reduce its column---it has to be $0$.
When computing cohomology, after reducing the coboundary matrix of
$k$-simplices, we can identify all the $(k+1)$-simplices that appear
as pivots $\low$ of some column and then set their columns to $0$ before
reducing them in dimension $k+1$.

\section{Algorithm}


As is often the case in applications, we assume our input data is spatially
partitioned.
We model this as a cover of the domain with sub-complexes.
To make this precise, we use the following notation.
Let $K$ be a simplicial complex covered by simplicial sub-complexes $K_i$,
$K = \bigcup_i K_i$.
A simplex $\sigma$ is called \textit{interior to $K_i$}, if it belongs only to $K_i$
and no other sub-complex,
$\sigma \in K_i \setminus \left( \bigcup_{j \neq i} K_i \cap K_j \right)$.
The rest of the simplices, $\bigcup_{i\neq j} (K_i \cap K_j)$, are called \textit{shared}.
The intersection of simplicial complexes is a simplicial complex;
therefore, a simplex is shared if and only if all of its vertices are shared.
If at least one vertex of a simplex lies in the interior $K_i \setminus \left( \bigcup_{j \neq i} K_i \cap K_j \right)$,
then the simplex is interior.

The coboundary matrix $D_i$ of $K_i$ consists of two parts\footnote{Normally it would be denoted $D^{\perp}_i$ to distinguish it from
the boundary matrix, but we only work with cohomology in this
paper, so we drop the $\perp$ superscript everywhere for convenience.}.
The columns of the interior simplices in $D_i$ consist entirely of interior
simplices and, therefore, they are the same as in the full coboundary matrix of $K$.
The columns of shared simplices in $D_i$ are subsets of the full columns in $D$;
the rows of simplices outside $K_i$ are missing.

After the initial local reduction of matrices $D_i$, we assemble the full matrix
$D$, partitioned among the processors.
Let $N$ be the number of columns of $D$, $p$ be the number of processors.
We make processor $i$ responsible for the column and row range $[s_{i},
s_{i+1})$, where the sequence
\[
    -\infty = s_{-1} < s_0 < s_1 < \cdots < s_{p-1} < s_p = \infty
\]
is computed to split the number of columns between processors approximately
evenly. We use a sample sort to identify the splitters.
Finding a rank that is responsible for value $v$ is done
by binary search, as expressed in \cref{alg:rank_by_value}.

\begin{algorithm}
\caption{Obtain rank by value.}
\label{alg:rank_by_value}
\begin{algorithmic}[1]
    \Function{RankByValue}{$v$}
        \State{$r \gets \mbox{first $i$ such that $s_{i-1} \leq v < s_i$}$}
        \State{\Comment{$r$ is found by binary search}}
        \State{\Return{$r$}}
    \EndFunction
\end{algorithmic}
\end{algorithm}

\begin{algorithm}
\caption{Full Algorithm}
\label{alg:full_algo}
\begin{algorithmic}[1]
    \State {$R_i \gets \Call{ReduceLocal}{K_i}$}
    \State {$[s_{-1}, \dots, s_p] \gets \Call{SampleSort}{\mbox{all simplex values}}$}
    \State {$\rfin \gets \Call{RedistributeColumns}{R_i, \,[s_{-1}, \dots, s_p]}$} 
    \State {$\rfin \gets \Call{Reduce}{\rfin}$}

    \State{$\mathtt{all\_done} \gets \mathtt{False}$}
    \State{$\mathtt{round} \gets 1$}
    \State {$\updated \gets \emptyset$}

    \For{$\mathtt{dim} = 0..\mathtt{max\_dim}$}
        \While{not \texttt{all\_done}}
            \State {\Call{SendColumns}{$\mathtt{dim}$, $\updated$, $\mathtt{round}$}}
            \State {$\texttt{all\_done} \gets \mbox{none of the ranks sent a column}$}
            \State {$\updated \gets \Call{ReceiveColumns}{}$}
            \State{$\mathtt{round} \gets \mathtt{round} + 1$}
        \EndWhile
        \State {\Call{ClearColumns}{$\mathtt{dim}$}}
    \EndFor
\end{algorithmic}
\end{algorithm}

The overall algorithm, expressed in pseudocode in \cref{alg:full_algo}, consists of three parts:
\begin{enumerate}
    \item
        Local reduction and sparsification. We reduce each $D_i$
        on a separate processor and then sparsify it, as discussed in \cref{sec:local,sec:sparsification}.
        At this phase, rank $i$ has the columns of $D$ that correspond to
        simplices in cover sub-complex $K_i$ (domain partitioning).
        The columns of shared simplices are incomplete.
    \item
        Rearranging matrix across ranks in filtration order. After determining
        the splitters $s_i$ via a sample sort, the processors exchange their
        domain partitioned columns, to partition them by function value. Rank
        $i$ gathers a contiguous chunk of columns of $D$ whose \textbf{column
        values} are in $\rseg{i}$.  All columns are now complete (they include
        both interior and shared simplices). Each rank runs a local reduction on
        its own chunk once.
    \item
        Global reduction. At this step, rank $i$ becomes responsible for the
        columns whose $\low$ belongs to segment $\rseg{i}$.  In a reduction
        loop, each rank
        \begin{enumerate*}
            \item receives columns sent to it,
            \item runs a local reduction using these received columns,
            \item sends columns whose $\low$ is not in $\rseg{i}$ to the rank
                responsible for it. The loop finishes when all columns are
                reduced.
        \end{enumerate*}
\end{enumerate}


\subsection{Local Reduction: Matrix Structure}
\label{sec:local}
\begin{figure}
    \centering
    \includegraphics{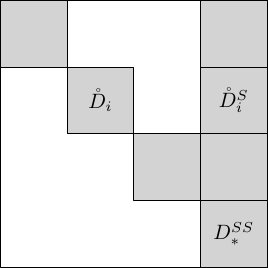}
    \caption{The overall structure of the coboundary matrix $D$. Non-zero blocks are
             highlighted in gray, and different blocks of submatrix $D_i$ are labeled.
             $D_*^{SS}$ contains rows and columns from $D_i^{SS}$ for all $i$. This
             coboundary matrix structure is the same as the boundary matrix structure of
             the blowup complex used in \cite{lewis2015parallel}.}
    \label{fig:matrix-structure}
\end{figure}

The initial local computations of our algorithm are as in \cite{lewis2015parallel}.
We denote by $\dint{i}$ the submatrix of $D_i$ formed by the columns that correspond to the interior simplices.
We denote by $\dsp$ the submatrix of $D_i$ formed by the columns that correspond to shared simplices.
The coboundary of a shared simplex can contain both shared and interior simplices.
We split $\dsp$ further into the submatrix $\dspi$ formed by the rows
of interior simplices and $\dsps$ formed by the rows of shared simplices.
See \cref{fig:matrix-structure}.
The order of columns and rows inside each $\dint{i}$, $\dspi$ and $\dsps$
is the same as in $D_i$.

\begin{figure}
    \centering
    \includegraphics{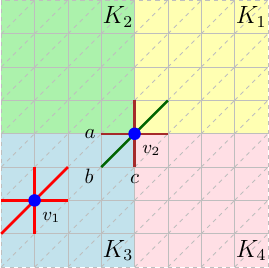}
    \caption{A triangulation of a 2-dimensional grid, partitioned among four
    processes. The four cover sub-complexes are shown in four
    colors. Coboundaries of interior ($v_1$) and shared ($v_2$) simplices are
    illustrated.}
    \label{fig:split_1}
\end{figure}

An example is given in \cref{fig:split_1}. The global domain is split
between 4 ranks.
We use Freudenthal triangulation to create a simplicial complex $K$
on the regular grid (dashed lines), covered by the four sub-complexes $K_1,
\ldots, K_4$.
The coboundary of an interior vertex $v_1$ consists of 
the 6 highlighted edges, all interior.
The part of the coboundary of a shared vertex $v_2$ that belongs to sub-complex
$K_3$, and therefore to the coboundary matrix $D_3$ on rank 3, consists of two
shared edges $\{a, v_2\}$ and $\{c, v_2\}$
and one interior edge $\{b, v_2\}$. This example also illustrates
why none of the ranks have complete columns of $D$ for the shared
simplices.

Crucially, only the rows of simplices interior to $K_i$ have non-zero entries in $\dint{i}$.
In other words, each rank has complete columns of the global coboundary matrix
$D$ that correspond to its interior simplices.
Therefore, rank $i$ can locally reduce matrix $\dint{i}$.
However, there is no way to guarantee (yet) that a local pair
$(\sigma, \tau)$ identified from the reduced $\dint{i}$ is a true global pair.

\cref{alg:reduce_init} contains pseudocode for the local part. Since we do not run
any reduction operations on the columns of $\dsp$, and all further parts
of the algorithm operate on matrix $R$ which was initially $D$, in the pseudocode
we directly write the columns of matrices $\dspi$ and $\dsps$
into the corresponding matrices $\rspi{i}$, $\rsps{i}$.

\begin{algorithm}
\caption{Local Reduction.}
\label{alg:reduce_init}
\begin{algorithmic}[1]
    \Function{ReduceLocal}{$K_i$}
        \State{$D_i \gets \mbox{coboundary matrix of $K_i$}$}
        \State{$\dint{i} \gets \mbox{empty matrix}$}
        \State{$\dsp \gets \mbox{empty matrix}$}
        \ForAll{columns $\col{D_i}{\sigma}$ of $D_i$}
            \If{$\sigma$ is interior}
                \State{append $\col{D_i}{\sigma}$ to $\dint{i}$}
            \Else
                \State{append empty columns to $\rspi{i}$, $\rsps{i}$}
                \ForAll{entries $\tau$ of $\col{D_i}$}
                    \If{$\tau$ is interior}
                        \State{append $\tau$ to the last column of $\rspi{i}$}
                    \Else
                        \State{append $\tau$ to the last column of $\rsps{i}$}
                    \EndIf
                \EndFor
            \EndIf
        \EndFor
        \State{$\rint{i} \gets \Call{ReduceMatrixELZ}{\dint{i}}$}
    \EndFunction
\end{algorithmic}
\end{algorithm}

\subsection{Local Reduction: Ultrasparsification}
\label{sec:sparsification}
Ultrasparsification is a procedure described in
\cite{lewis2015parallel} that allowed its authors to lower both the space and
time complexity of the global hierarchical reduction for spatially partitioned
data.  The idea is to zero out all but one elements of the interior columns via
row operations. The validity of this operation, stated in the following lemma,
follows from \cite{vineyards}.

\begin{lemma}
    The persistence pairing determined by $R$ does not change,
    if in the reduction process we also perform row operations
    $\row{R}{i} \gets \row{R}{i} + \row{R}{j}$ provided
    that we add rows bottom-up, $j > i$.
\end{lemma}

When we are done reducing a column $\col{R}{k}$, we can add the row $\low(\col{R}{k})$
to all other rows $j$ such that $R[j, k] \neq 0$. As a result, the only non-zero entry remaining
in the column is the pivot. In a sequential
algorithm, this would add extra work, but it would produce the same persistence
pairing.
In the distributed setting, this procedure both reduces the size of the columns
that we send between ranks and makes column addition faster.

In our setting, we can perform this operation locally,
because each rank $i$ has complete rows of matrix $D$ that correspond
to the simplices of $K_i$. Since we split the local matrix into
parts, we must apply row additions simultaneously to $\rint{i}$ and $\rspi{i}$.
Note that $\rsps{i}$, which is shared among multiple ranks, remains untouched.
A convenient optimization is to perform ultrasparsification itself bottom-up,
starting with the column of $\rint{i}$ whose $\low$ is maximal. Then we do not actually
need to do any computations on $\rint{i}$, because we maintain
the invariant: when we add row $\ell$ to row $t$ above it,
the only non-zero element in row $\ell$ is the lowest one
of the column that we are processing, so we
can we can just remove all elements from the column of $\rint{i}$
except for the lowest one. We must,
however, perform a row addition on $\rspi{i}$.
So, while the interior columns become ultrasparse,
the columns of shared simplices can actually become denser.

We perform one more operation to sparsify the columns of $\rspi{i}$.
In the global matrix $R$, the columns of $\rint{i}$ and $\rspi{i}$ are
interleaved, following the reverse filtration order.
Since we can always perform column operations,
we can add the ultrasparse column $\col{\rint{i}}{t}$
to every column of $\rspi{i}$ that succeeds it in the reverse filtration
order and contains $x = \low(\col{\rint{i}}{t})$. This operation removes
$x$ from the column $\rspi{i}$.

\begin{figure*}
    \centering
    \includegraphics[width=\textwidth]{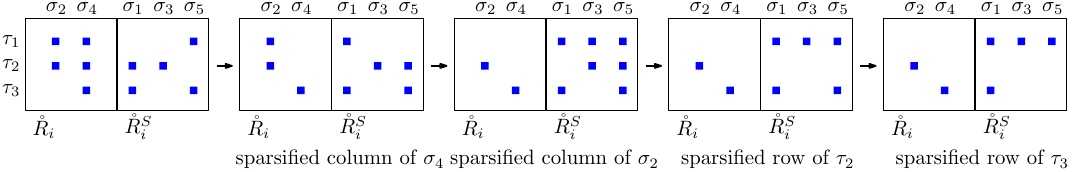}
    \caption{Illustration of the ultrasparsification procedure.}
    \label{fig:sparsify_1}
\end{figure*}

\cref{fig:sparsify_1} illustrates the overall procedure.
We write $\alpha \prec \beta$ to indicate that in the global matrix $R$ the column of $\alpha$ precedes the column of $\beta$.
Global order of simplices in matrix $R$ corresponds
to simplex indices: $\sigma_{i} \prec \sigma_{i+1}$ for $i = 1,\dots,4$. 
We start the ultrasparsification at column $\col{\rint{i}}{\sigma_4}$
because its pivot is the lowest. We add the row of $\tau_3$ to the rows of $\tau_2$ and $\tau_1$ to remove
all non-zero entries from the column. We must perform the same additions on the rows of $\rspi{i}$. Then we sparsify the column
$\col{\rint{i}}{\sigma_2}$. After the columns of $\rint{i}$ are ultrasparse, we can perform column operations
to sparsify the rows. We add column $\col{\rint{i}}{\sigma_2}$ to
all columns of $\rspi{i}$ that are to the right from it and contain $1$ in row $\tau_2$
(in the figure, we remove $\tau_2$ from both columns $\col{\rspi{i}}{\sigma_3}$ and $\col{\rspi{i}}{\sigma_5}$).
When we sparsify the row of $\tau_3$, we cannot add $\col{\rint{i}}{\sigma_4}$ to $\col{\rspi{i}}{\sigma_1}$,
because $\sigma_4$ is to the left from $\sigma_1$, so we can only remove one entry by adding $\col{\rint{i}}{\sigma_4}$
to $\col{\rspi{i}}{\sigma_5}$.

Pseudocode for this procedure is given in \cref{alg:ultrasparsify}. Here we slightly
abuse the notation: since we work over $\ZZ/2\ZZ$, we treat columns as sets. 

\begin{algorithm}
\caption{Ultrasparsification.}
\label{alg:ultrasparsify}
\begin{algorithmic}[1]
    \Function{Sparsify}{$\rint{i}, \rspi{i}$}
        \State{$L \gets \mbox{sorted }\{ \low(\col{\rint{i}}{c}) | \mbox {non-zero columns $c$ of }\rint{i} \} $}
        \For{$\ell \in L$}
            \State{\Comment{$t$ the column of $\rint{i}$ whose $\low$ is $\ell$}}
            \State{$ t \gets \piv{\ell}$}
            \ForAll{$e \in \col{R}{t}$ except $\ell$}
                \State{$\row{{\rspi{i}}}{e} \gets \row{\rspi{i}}{e} + \row{\rspi{i}}{\ell}$}
            \EndFor
            \State{\Comment{the only entry left is $\low$}}
            \State{$\col{\rint{i}}{t} \gets \{\ell\}$}
        \EndFor
        \ForAll{column $\col{\rspi{i}}{j}$ of $\rspi{i}$}
            \ForAll{entry $e \in \col{\rspi{i}}{j}$}
        \If{$\exists p\mbox{ s.t. } \low(\col{\rint{i}}{t}) = e \mbox{ and $p \prec j$}$}
                \State{\Comment{removes $e$ from $\col{\rspi{i}}{j}$}}
                \State{$\col{\rspi{i}}{j} \gets \col{\rspi{i}}{j} + \col{\rint{i}}{p}$}
            \EndIf
            \EndFor
        \EndFor
    \EndFunction
\end{algorithmic}
\end{algorithm}

\subsection{Rearranging}
\label{sec:rearrange}
When we compute local coboundary matrix $D_i$, we only
have access to the values of simplices in $K_i$.
Suppose
\[
    K_i = \{ \sigma_1^i, \dots, \sigma_{N_i}^i \},
\]
where we list simplices in the reverse filtration order:
\[
    v_1^{(i)} > \cdots > v_{N_i}^{(i)},\mbox{ where } v_t^{(i)} = \val(\sigma_t^i).
\]
In the compressed column representation,
if a column of $D_i$ is $[a_1, a_2, \dots]$,
$0 \leq a_1 < a_2 < \cdots$, it means
that it contains simplices $\sigma_{a_1}^i$, $\sigma_{a_2}^i$, etc.
In order to perform the global reduction, we must
use consistent global indexing of simplices.
If we order all simplices following the filtration,
\[
    K = \{ \sigma_1, \dots, \sigma_{N} \}, \mbox{ and } v_1 > \cdots > v_{N},\mbox{ where } v_t = \val(\sigma_t),
\]
then, since $K = \bigcup K_i$, every simplex $\sigma_t^i$ in the local order of $K_i$ maps to some
$\sigma_{t'}$ in the global order,
\[
    \forall i \in \{1, \dots, p \}, ~ \forall t \in \{ 1, 2, \dots, N_i \}, ~
    \exists !~ t' \mbox{ such that } \sigma_t^{(i)} = \sigma_{t'}.
\]
This generates a monotonically increasing map,
\[
    \rindx_i \colon \{ 1, \dots, N_i \} \to \{ 1, \dots, N \},\quad t \mapsto t'.
\]
To go from matrices $R_i$ to a global matrix $R$, we must translate all entries
using the map $\rindx_i$.

This approach is not feasible in practice: it requires
having the sorted values, $\{v_1, \ldots, v_N\}$, of the entire domain
in the memory of each rank. This is impossible for large datasets,
especially when we take into account that the reduction algorithm
is itself memory-intensive.

An alternative strategy
is to avoid integral indices and to store the values themselves:
instead of having columns of $\rfin$ of the form $[a_1, a_2, \dots]$,
where $a_t$ is the global index of $\sigma_t \in K$,
we will represent the columns as sorted arrays of values $v_t$ directly.
The complexity of column addition does not change; it is the same
symmetric difference.

\begin{algorithm}
\caption{Rearranging Matrix.}
\label{alg:rearrange}
\begin{algorithmic}[1]
    \Function{RedistributeColumns}{$R_i$}
        \ForAll{columns $\col{R_i}{\sigma}$ of $R_i$}
            \State{$r \gets $\Call{RankByValue}{$\low(\col{R_i}{\sigma})$}}
            \State{convert $\col{R_i}{\sigma}$ from indices to values}
            \State{send $\col{R_i}{\sigma}$ to $r$}
        \EndFor
        \ForAll{incoming columns $\col{R_k}{\sigma}$}
        \State{\Comment{merge $\col{R_k}{\sigma}$ into $\rfin$}}
            \If{$\rfin$ already has column for $\sigma$}
                \State{$\col{\rfin}{\sigma} \gets \col{R_k}{\sigma} \cup \col{\rfin}{\sigma} $}
            \Else
                \State{$\col{\rfin}{\sigma} \gets \col{R_k}{\sigma}$}
            \EndIf
        \EndFor
        \State{$\rfin \gets \Call{ReduceMatrixELZ}{\rfin}$}
    \EndFunction
\end{algorithmic}
\end{algorithm}

Many columns are zeroed by local reduction,
so we choose to represent the global matrix not as an \texttt{std::vector}
of \texttt{std::vector}s, but as an \texttt{std::map} that
uses column values as keys. The columns themselves are still stored
as \texttt{std::vector}s of values. The fact that a \texttt{map}
is ordered makes iterating over columns in the filtration order
easy. If during the global reduction a column becomes zero,
we remove it from the map entirely, to avoid storing empty vectors.

We must eliminate all collisions between columns in the whole matrix.
Therefore, it is convenient to rearrange columns among ranks according
to the values $\low$ of their pivots (this idea was proposed in \cite{bauer2014distributed}).
We cannot determine this assignment immediately after the local reduction step:
ranks do not have complete columns for shared simplices.
Before we go into the global reduction loop,
we rearrange the matrix according to column values: the column of $\sigma$
is sent to rank $i$ such that $\val(\sigma) \in \rseg{i}$. When a rank
receives columns, it needs to merge different parts of a column together.
The ultrasparsification procedure does not create any conflicts,
because each rank was operating on the rows that correspond to
its interior simplices. After we assemble a contiguous chunk of columns $\rfin$
on each rank, we run a local reduction on this chunk again.
The pseudocode for this part is in \cref{alg:rearrange}.
To avoid additional verbosity, in this algorithm we do not explicitly
state that columns of the local matrix $R_i$ are actually split between $\rint{i}$, $\rspi{i}$
and $\rsps{i}$.

\subsection{Global reduction loop}

The main reduction loop is conceptually simple: rank $i$ receives columns
whose $\low$ belongs to the range $\rseg{i}$ and reduces them
until there are no more collisions among local columns.
Columns that were reduced to zero are removed from the map;
non-zero columns are sent to the rank responsible for their new $\low$.
Note that when we reduce a column, its $\low$ only gets smaller (moves up
in the matrix). Therefore we only send columns from rank $i$ to rank $j < i$.

There is one exception: in the first iteration of the loop, we must send all columns
to new ranks. Recall that after rearranging column $\col{R}{\sigma}$ is on the
rank that corresponds to $\val(\sigma)$, not to the value of its $\low$. 
In all other iterations, we must send every column whose
$\low$ moved out of $\rseg{i}$ after we received columns and 
reduced them. These columns are contained in the $\mathtt{updated}$
argument in \cref{alg:send_columns}.

\begin{figure}
    \centering
    \includegraphics[width=\columnwidth]{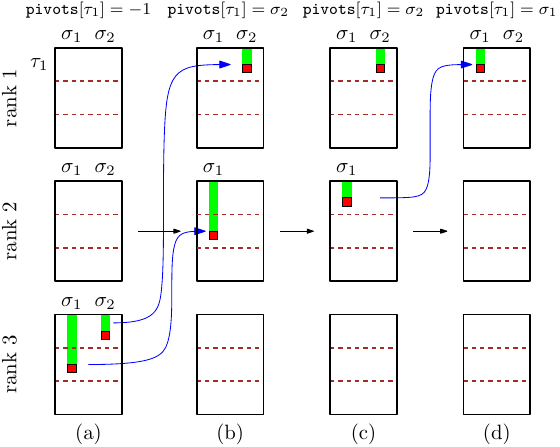}
    \caption{An example of the pivot to the right situation. (a) Rank 3 has
             $\col{R}{\sigma_1}$ and $\col{R}{\sigma_2}$. Rank 1 does not have
             any column whose $\low$ is $\tau_1$, so in its local pivots table
             the corresponding entry is $-1$. (b) Based on their $\low$ values,
             rank 3 sends the columns to ranks 2 and 1, respectively. (c) After
             the local reduction, on rank 1, $\col{R}{\sigma_2}$ has no
             collisions, so $\sigma_2$ remains a pivot for $\tau_1$. Rank 2
             reduces $\col{R}{\sigma_1}$ (via columns that are not shown) and
             its new $\low$ is now $\tau_1$. (d) Rank 2 sends
             $\col{R}{\sigma_1}$ to rank 1. Now, when rank 1 runs local
             reduction on $\col{R}{\sigma_1}$, its local pivots table points
             to $\col{R}{\sigma_2}$ which is \textit{to the right} from
             $\sigma_1$.}
    \label{fig:right_pivot}
\end{figure}

The local reduction procedure is slightly more complicated than the standard reduction.
In \cref{alg:elz}, when we reduce column $\col{R}{\sigma}$ and query
the pivots table for $\piv{\low(\col{R}{\sigma})}$, the column in the table
is always to the left. In the global reduction loop, we may receive columns that
lie to the left of us in the global order, and so the existing pivot may lie to
the right; see \cref{fig:right_pivot} for an illustration.
In this case, we mark the column we are reducing as the new pivot and switch to
reducing the column to the right that was marked as the pivot before.
This algorithm is closely related to the algorithm in
\cite{morozov2020towards}, where it is used for shared-memory parallelism.
Its adaptation to our setting makes it simpler and obviates the need for atomic
operations and complicated memory management.
The resulting pseudocode is in \cref{alg:local_fin}.

\begin{algorithm}
\caption{Sending columns in global reduction.}
\label{alg:send_columns}
\begin{algorithmic}[1]
    \Function{SendColumns}{$\mathtt{dim}$, $\updated$, $\mathtt{round}$}
        \If {$\mathtt{round} = 1$}
            \State{\Comment{first round: rearrange \textbf{all} columns by $\low$}}
            \ForAll{column $\col{\rfin}{\sigma}$ of $\rfin$ in dimension $\mathtt{dim}$}
                \State{$r \gets \Call{RankByValue}{\low(\col{\rfin}{\sigma})}$}
                \State{send $\col{\rfin}{\sigma}$ to rank $r$}
            \EndFor
        \Else
            \ForAll{column $\col{\rfin}{\sigma}$ where $\sigma \in \updated$}
                \State{\Comment{guaranteed: $\dim(\sigma) = \mathtt{dim}$}}
                \State{$r \gets \Call{RankByValue}{\low(\col{\rfin}{\sigma})}$}
                \State{send $\col{\rfin}{\sigma}$ to rank $r$}
            \EndFor
        \EndIf
    \EndFunction
\end{algorithmic}
\end{algorithm}

\begin{algorithm}
\caption{Receiving Columns in Distributed Reduction.}
\label{alg:local_fin}
\begin{algorithmic}[1]
    \Function{ReceiveColumns}{$C$} \Comment{$C$: received columns}
        \State {$\updated \gets \emptyset$}
        \State{$\sigma \gets \mbox{leftmost of all columns in }C$} \Comment{we start reduction from $\sigma$}
        \State{insert columns of $C$ into $\rfin$}
        \While{\textbf{true}}
            \If{$\sigma$ is not the last column in $\rfin$}
                \State{$\sigma_{next} \gets \mbox{next simplex after $\sigma$ in $\rfin$}$}
            \Else
                \State{$\sigma_{next} \gets \sigma$}
            \EndIf

            \While{$\col{\rfin}{\sigma} \neq 0$}
                \State {$\ell \gets \low(\col{\rfin}{\sigma})$}
                \If {$\piv{\ell} \prec \sigma$ and $\piv{\ell} \neq -1$}
                    \State \Comment {Pivot to the left, standard reduction}
                    \State{$\col{\rfin}{\sigma} \gets \col{\rfin}{\sigma} + \col{\rfin}{\piv{\ell}}$}
                \ElsIf {$\piv{\ell} \succ \sigma$}
                    \State \Comment {Pivot to the right, switch to reducing column $\piv{\ell}$}
                    \State{\Call{Swap}{$\piv{\ell}, \sigma$}}
                    \State{$\col{\rfin}{\sigma} \gets \col{\rfin}{\sigma} + \col{\rfin}{\piv{\ell}}$}
                \Else
                    \State \Comment {We saw that column and reduced it}
                    \State{\textbf{break}}
                \EndIf
            \EndWhile

            \If{$\col{\rfin}{\sigma} \neq 0$}
                \State{$\piv{\low(\col{\rfin}{\sigma})} \gets \sigma$}
                \If {$\low(\col{\rfin}{\sigma}) \notin \rseg{i}$}
                    \State \Comment {$\col{\rfin}{\sigma}$ no longer belongs }
                    \State \Comment {to current rank $i$, mark it to send further}
                    \State{insert $\low(\col{\rfin}{\sigma})$ into $\updated$}
                \EndIf
            \Else
                \State{erase column $\col{\rfin}{\sigma}$}
            \EndIf

            \If{$\sigma_{next} \neq \sigma$}
                \State{$\sigma \gets \sigma_{next}$}
            \Else
                \State{\textbf{break}}
            \EndIf
        \EndWhile
        \State {\Return $\updated$}
    \EndFunction
\end{algorithmic}
\end{algorithm}


\section{Clearing}
\begin{algorithm}
\caption{Global Clearing.}
\label{alg:global_clear}
\begin{algorithmic}[1]
    \Function{ClearColumns}{$\mathtt{dim}$}
    \ForAll{non-zero columns $\col{\rfin}{\tau}$ of $\rfin$ in dimension $\mathtt{dim}$}
            \State {$\sigma \gets \low(\col{\rfin}{\tau})$} \Comment {Dimension of $\sigma$ is $\mathtt{dim} + 1$}
            \State {$r \gets \Call{RankByValue}{\sigma}$}
            \State {send $\sigma$ to $r$}
        \EndFor
        \ForAll{incoming $\sigma$}
            \State {erase column $\col{\rfin}{\sigma}$}
        \EndFor
    \EndFunction
\end{algorithmic}
\end{algorithm}

An important aforementioned optimization is clearing: zeroing out columns of
positive simplices without reducing them explicitly.  Applying it during the
initial local phase is straightforward.  If the row of some simplex $\sigma$
contains the lowest non-zero entry of at least one column, this will be true
until the end of the reduction (either there will be no collisions, or this
particular lowest one will be killed by adding another column \textit{with the
same $\low$}). Therefore, if the column of $\sigma$ happens to be at the same
rank, it can be zeroed out immediately.  We only need to make sure that we
perform reduction in the increasing dimension order. This can be achieved by
(stable) sorting the simplices in the filtration with respect to the dimension.

As for the global phase, we choose to explicitly split
the matrix into dimension components and perform clearing
after we finish each dimension. The reason is that before
we enter the global reduction loop (before $SendColumns$
is executed), the columns of the matrix in the next dimension
are assigned to ranks by their column value. This allows
us to easily identify the rank holding the column
that we are going to zero out and send the message to this rank only,
as in \cref{alg:global_clear}.

Our experiments show that, while the fraction of the columns zeroed by the
global clearing is small, this optimization is crucial in reducing the running
time, because those columns would require many more communication rounds to
explicitly reduce to zero.

\section{Experiments}
We compare our implementation, called \cadmus, with \dipha
\cite{bauer2014distributed}, the only publicly available distributed
implementation of persistence. Both codes use MPI; our code uses block-parallel
library \diy \cite{morozov2016block} for domain partitioning and exchange
between blocks.
Experiments were performed on the Perlmutter supercomputer (CPU Nodes) at the
National Energy Research Scientific Center (NERSC).
Every node has 2 AMD Epyc 7763 processors (64 CPU cores, 2.45GHz) and 512 GB of RAM.
Our code and \dipha were compiled with Clang 13 specifically optimized by the vendor
for AMD processors (AOCC programming environment).

For all the experiments, both \dipha and \cadmus use
\textit{cubical} complexes, not simplicial ones
described in the background.
The reduction algorithm remains the same, but
the basic blocks are cubes. This approach is much more
natural for functions on grids and leads to filtrations
of smaller size.

Timings for \dipha are reported according to their own
benchmark option. We only take the time for the reduction
algorithm itself.  Since we use cohomology, we also run \dipha for cohomology.

All dataset files were downloaded from P. Klacansky's ``Open scientific
visualization datasets'' repository \cite{klacdb}.  We used the following data sets.
\begin{enumerate}
    \item Woodbranch is a microCT scan of a branch of hazelnut.
    \item Magnetic reconnection dataset is from \cite{magrec}. It is a simulation of an
        interaction between magnetic fields that were concentrated
        in two different regions and were pointing in two different directions;
        at the boundary between the regions, magnetic lines
        change connectivity.
    \item Isotropic pressure \cite{ip}. The scalar function here is enstrophy.
        Its high values correspond to regions where the magnitude of the vorticity 
        vector is large.
    \item Truss \cite{truss} is a simulated CT scan of a truss with some mechanical defects.
    \item Temperature field of the rotating stratified turbulence simulation \cite{rotstrat}.
\end{enumerate}

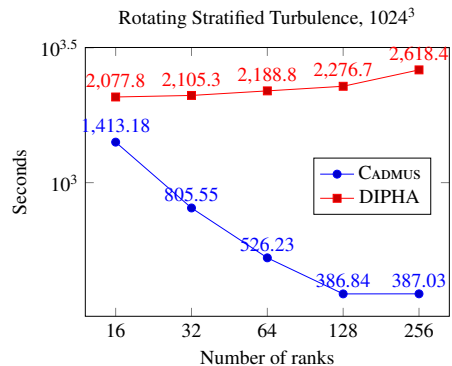
\begin{figure}[]
\centering
\begin{minipage}{0.45\textwidth}
\centering
\begin{tikzpicture}[scale=0.75]
\begin{axis}[
                title={Rotating Stratified Turbulence, $1024^3$},
                ymode=log,
                symbolic x coords={16,32,64,128,256},
                xtick={16,32,64,128,256},
                xticklabels={16,32,64,128,256},
                height=2.5in,
                width=.99\textwidth,
                xlabel={Number of ranks},
                ylabel={Seconds},
                nodes near coords,
                point meta=rawy,
                nodes near coords/.style={color=black},
                legend style={at={(0.95,0.59)}},
                legend columns=1,
            ]
    \addplot    +[] table[x=ranks, y=cadmus, col sep = tab]    {data/rotstrat_temperature_1024.csv}; \addlegendentry{\cadmus}
    \addplot    +[] table[x=ranks, y=dipha_coh, col sep = tab] {data/rotstrat_temperature_1024.csv}; \addlegendentry{\dipha}
\end{axis}
\end{tikzpicture}
    \caption{Strong scaling on the temperature field of rotating stratified turbulence simulation data set of size $1024^3$. Reduction time only.}
\label{fig:rotstrat_1024}
\end{minipage}
\end{figure}

The strong scaling plots in \cref{fig:wood_512,fig:mag_rec_512,fig:ip_512} are for $512^3$ data sets.
In all these examples \cadmus scales significantly better than \dipha and usually outperforms it. Same holds
for \cref{fig:ip_1024,fig:rotstrat_1024}, where the size is $1024^3$.

\begin{figure}[]
\centering
\begin{minipage}{0.45\textwidth}
\centering
\begin{tikzpicture}[scale=0.75]
\begin{axis}[
                title={Woodbranch, $512^3$},
                ymode=log,
                symbolic x coords={16,32,64,128},
                xtick={16,32,64,128},
                xticklabels={16,32,64,128},
                height=2.5in,
                width=.99\textwidth,
                xlabel={Number of ranks},
                ylabel={Seconds},
                nodes near coords,
                point meta=rawy,
                nodes near coords/.style={color=black},
                legend style={at={(0.42,0.18)},anchor=east},
                legend columns=1,
            ]
    \addplot    +[] table[x=ranks, y=cadmus, col sep = tab]    {data/woodbranch_512.csv}; \addlegendentry{\cadmus}
    \addplot    +[] table[x=ranks, y=dipha_coh, col sep = tab] {data/woodbranch_512.csv}; \addlegendentry{\dipha}
\end{axis}
\end{tikzpicture}
    \caption{Strong scaling on a 3D scan data (Woodbranch) of size $512^3$. Reduction time only.}
\label{fig:wood_512}
\end{minipage}
\end{figure}
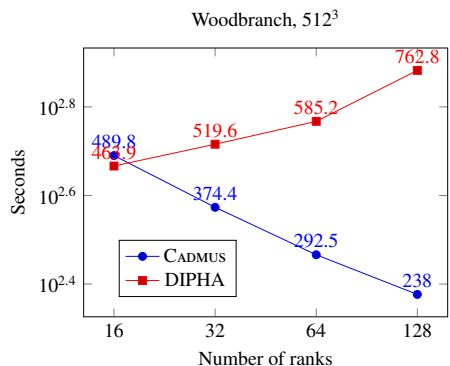
 
\begin{figure}[]
\centering
\begin{minipage}{0.45\textwidth}
\centering
\begin{tikzpicture}[scale=0.75]
\begin{axis}[
                title={Magnetic Reconnection, $512^3$},
                ymode=log,
                symbolic x coords={16,32,64,128},
                xtick={16,32,64,128},
                xticklabels={16,32,64,128},
                height=2.5in,
                width=.99\textwidth,
                xlabel={Number of ranks},
                ylabel={Seconds},
                nodes near coords,
                point meta=rawy,
                nodes near coords/.style={color=black},
                legend style={at={(0.42,0.18)},anchor=east},
                legend columns=1,
            ]
    \addplot    +[] table[x=ranks, y=cadmus, col sep = tab]    {data/mag_reconn_neg_512.csv}; \addlegendentry{\cadmus}
    \addplot    +[] table[x=ranks, y=dipha_coh, col sep = tab] {data/mag_reconn_neg_512.csv}; \addlegendentry{\dipha}
\end{axis}
\end{tikzpicture}
    \caption{Strong scaling on magnetic reconnection data set of size $512^3$. Reduction time only.}
\label{fig:mag_rec_512}
\end{minipage}
\end{figure}
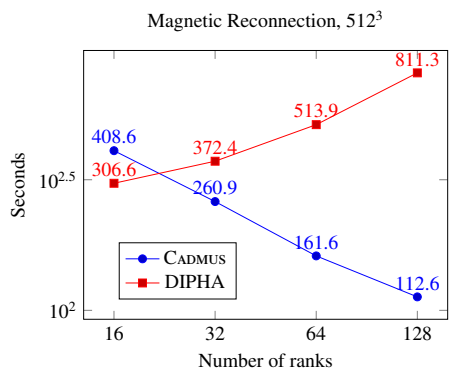

\begin{figure}[]
\centering
\begin{minipage}{0.45\textwidth}
\centering
\begin{tikzpicture}
\begin{axis}[
                title={Isotropic Pressure, $512^3$},
                ymode=log,
                symbolic x coords={16,32,64,128},
                xtick={16,32,64,128},
                xticklabels={16,32,64,128},
                height=2.5in,
                width=.99\textwidth,
                xlabel={Number of ranks},
                ylabel={Seconds},
                nodes near coords,
                point meta=rawy,
                nodes near coords/.style={color=black},
                legend style={at={(0.39,0.18)},anchor=east},
                legend columns=1,
            ]
    \addplot    +[] table[x=ranks, y=cadmus, col sep = tab]    {data/isotropic_pressure_512.csv}; \addlegendentry{\cadmus}
    \addplot    +[] table[x=ranks, y=dipha_coh, col sep = tab] {data/isotropic_pressure_512.csv}; \addlegendentry{\dipha}
\end{axis}
\end{tikzpicture}
    \caption{Strong scaling on isotropic pressure data set of size $512^3$. Reduction time only.}
\label{fig:ip_512}
\end{minipage}
\end{figure}

\begin{figure}[]
\centering
\begin{minipage}{0.45\textwidth}
\centering
\begin{tikzpicture}[scale=0.75]
\begin{axis}[
                title={Isotropic Pressure, $1024^3$},
                ymode=log,
                symbolic x coords={32,64,128,256},
                xtick={32,64,128,256},
                xticklabels={32,64,128,256},
                height=2.5in,
                width=.99\textwidth,
                xlabel={Number of ranks},
                ylabel={Seconds},
                nodes near coords,
                point meta=rawy,
                nodes near coords/.style={color=black},
                legend style={at={(0.95,0.59)}},
                legend columns=1,
            ]
    \addplot    +[] table[x=ranks, y=cadmus, col sep = tab]    {data/isotropic_pressure_1024.csv}; \addlegendentry{\cadmus}
    \addplot    +[] table[x=ranks, y=dipha_coh, col sep = tab] {data/isotropic_pressure_1024.csv}; \addlegendentry{\dipha}
\end{axis}
\end{tikzpicture}
    \caption{Strong scaling on isotropic pressure data set of size $1024^3$. Reduction time only.}
\label{fig:ip_1024}
\end{minipage}
\end{figure}

\begin{figure}[]
\centering
\begin{minipage}{0.45\textwidth}
\centering
\begin{tikzpicture}[scale=0.75]
\begin{axis}[
                title={Synthetic Truss, $600^3$},
                ymode=log,
                symbolic x coords={16,32,64,128},
                xtick={16,32,64,128},
                xticklabels={16,32,64,128},
                height=2.5in,
                width=.99\textwidth,
                xlabel={Number of ranks},
                ylabel={Seconds},
                nodes near coords,
                point meta=rawy,
                nodes near coords/.style={color=black},
                legend style={at={(0.42,0.82)},anchor=east},
                legend columns=1,
            ]
    \addplot    +[] table[x=ranks, y=cadmus, col sep = tab]    {data/truss_600.csv}; \addlegendentry{\cadmus}
    \addplot    +[] table[x=ranks, y=dipha_coh, col sep = tab] {data/truss_600.csv}; \addlegendentry{\dipha}
\end{axis}
\end{tikzpicture}
    \caption{Strong scaling on Synthetic Truss data set of size $600^3$. Reduction time only.}
\label{fig:truss_600}
\end{minipage}
\end{figure}
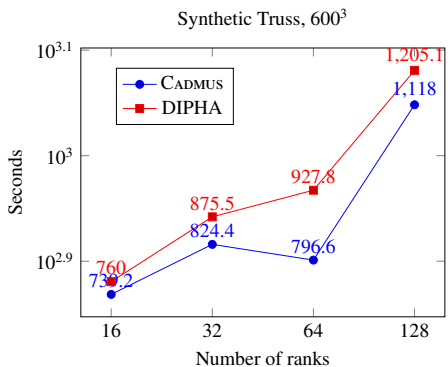

\begin{figure}[]
    \centering
    \includegraphics[width=0.8\columnwidth]{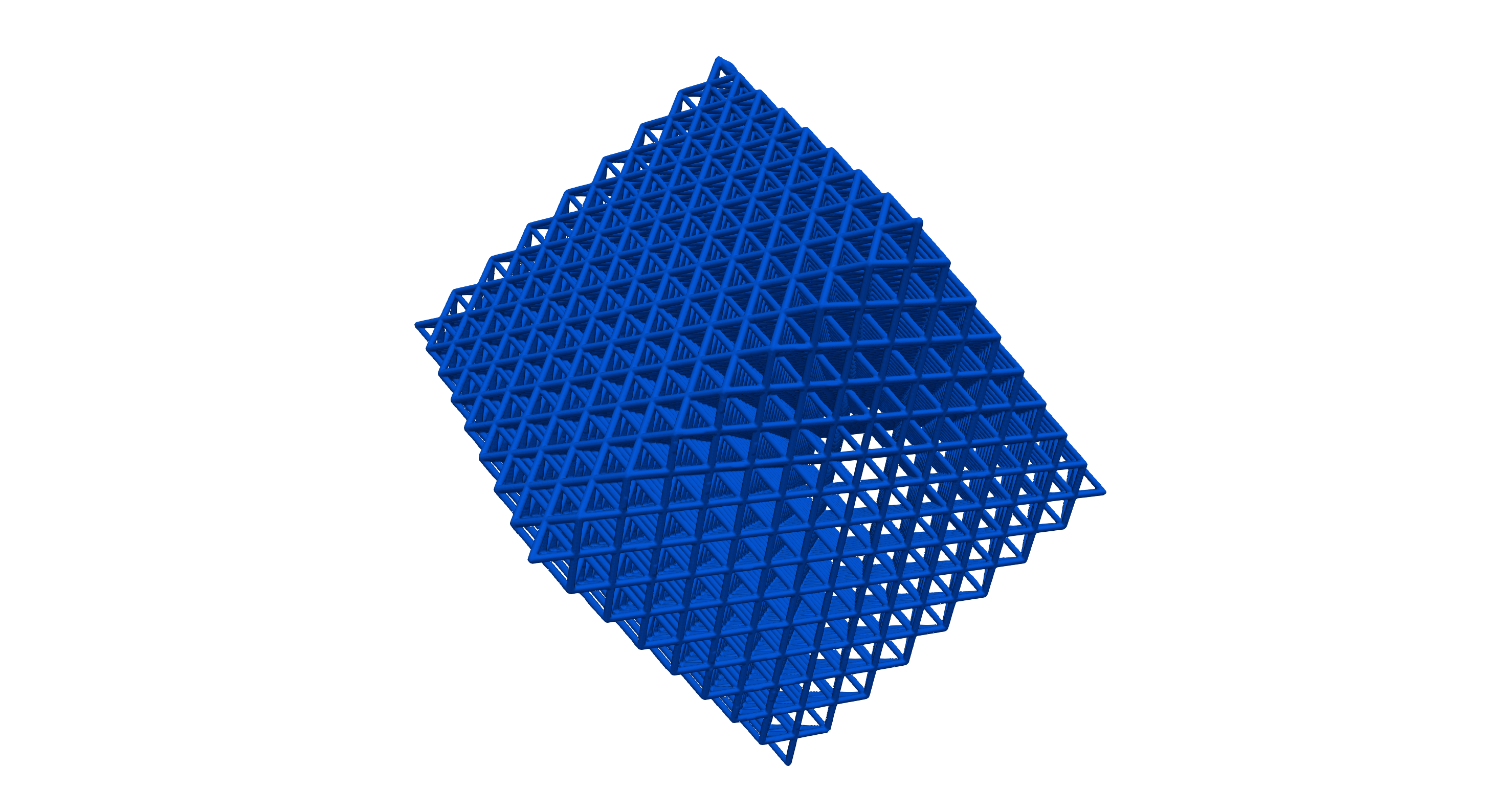}
    \caption{Visualization of Synthetic Truss dataset.}
    \label{fig:truss_vis}
\end{figure}

\begin{figure}
    \centering
    \includegraphics[width=0.7\columnwidth]{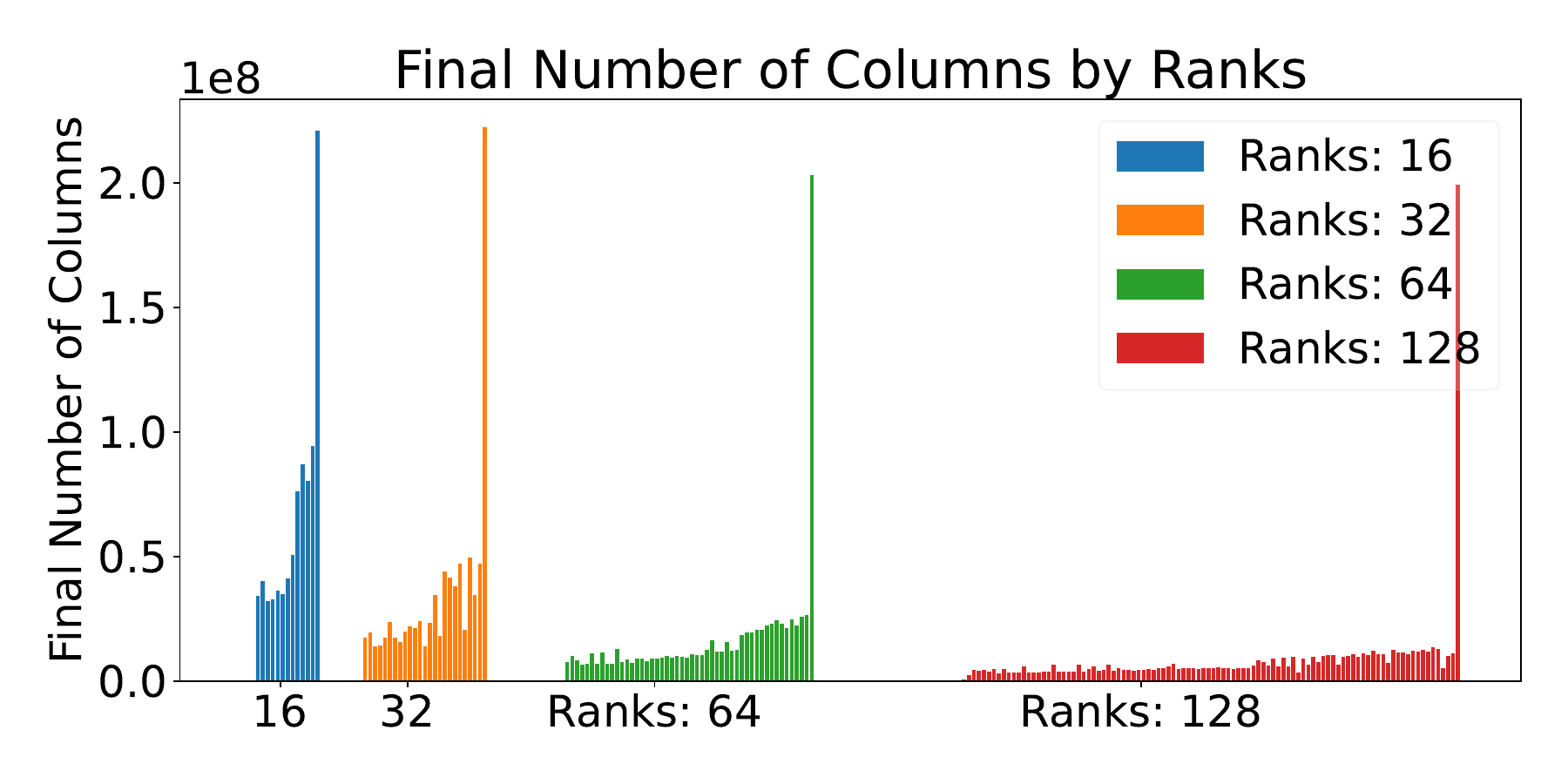}
    \caption{Histogram of the final number of columns on each rank for different number of ranks for Synthetic Truss dataset. One rank dominates all the others.}
    \label{fig:truss_col_stats}
\end{figure}

\begin{figure}
    \centering
    \includegraphics[width=0.7\columnwidth]{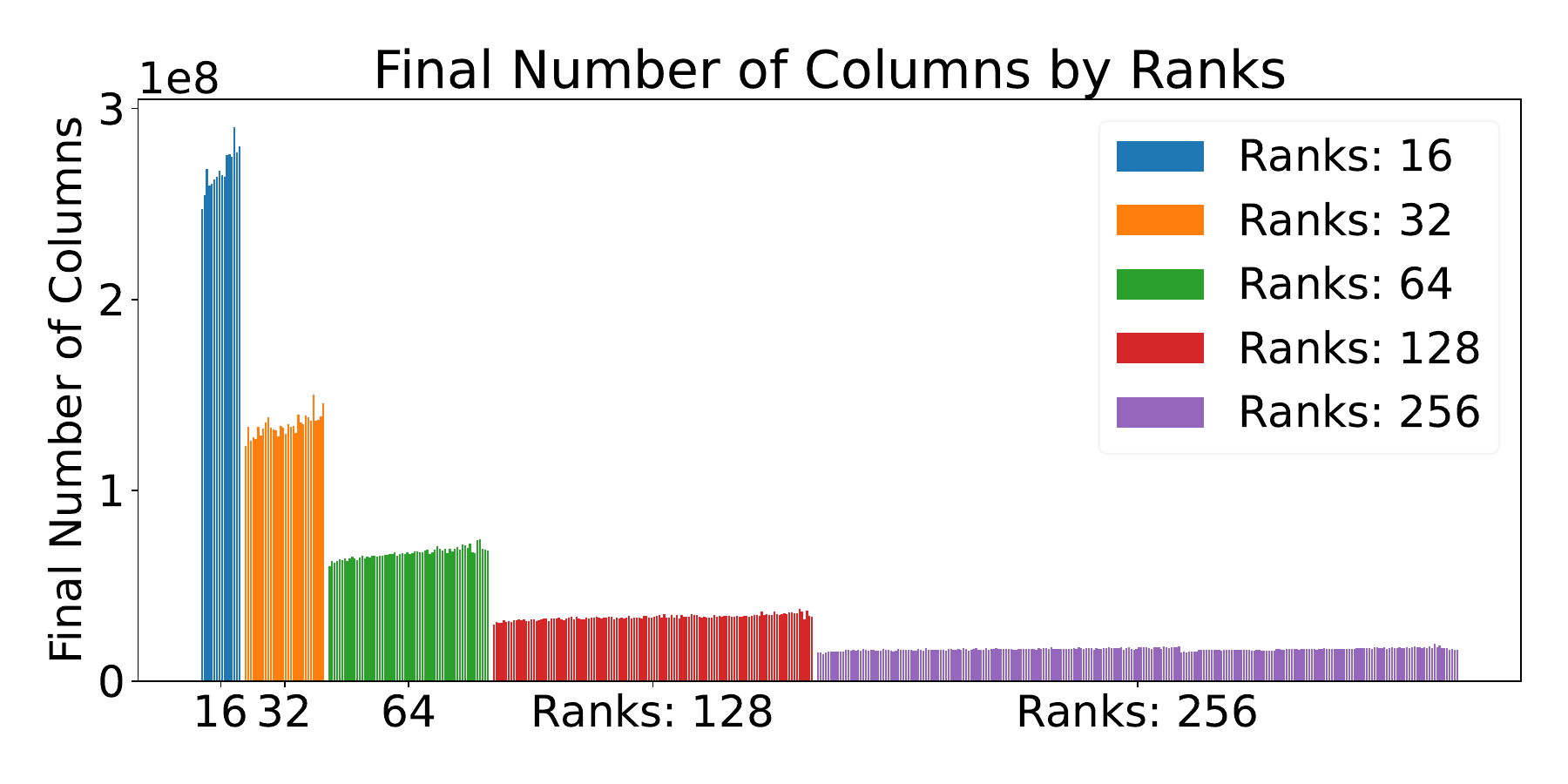}
    \caption{Histogram of the final number of columns on each rank for different number of ranks for rotating stratified turbulence dataset of size $1024^3$. The columns are equally distributed.}
    \label{fig:rotstrat_col_stats}
\end{figure}

\begin{figure}
    \centering
    \includegraphics[width=0.7\columnwidth]{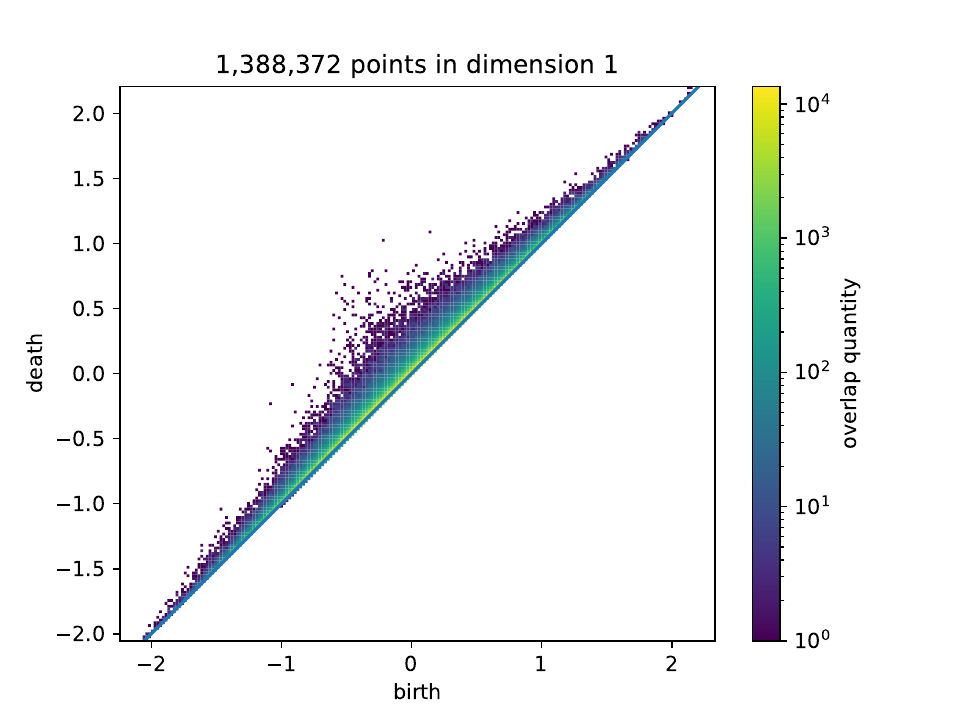}
    \caption{Persistence diagram of the rotating stratified turbulence data set in dimension 1.}
    \label{fig:rotstrat_dgm}
\end{figure}

\begin{figure}
    \centering
    \includegraphics[width=0.7\columnwidth]{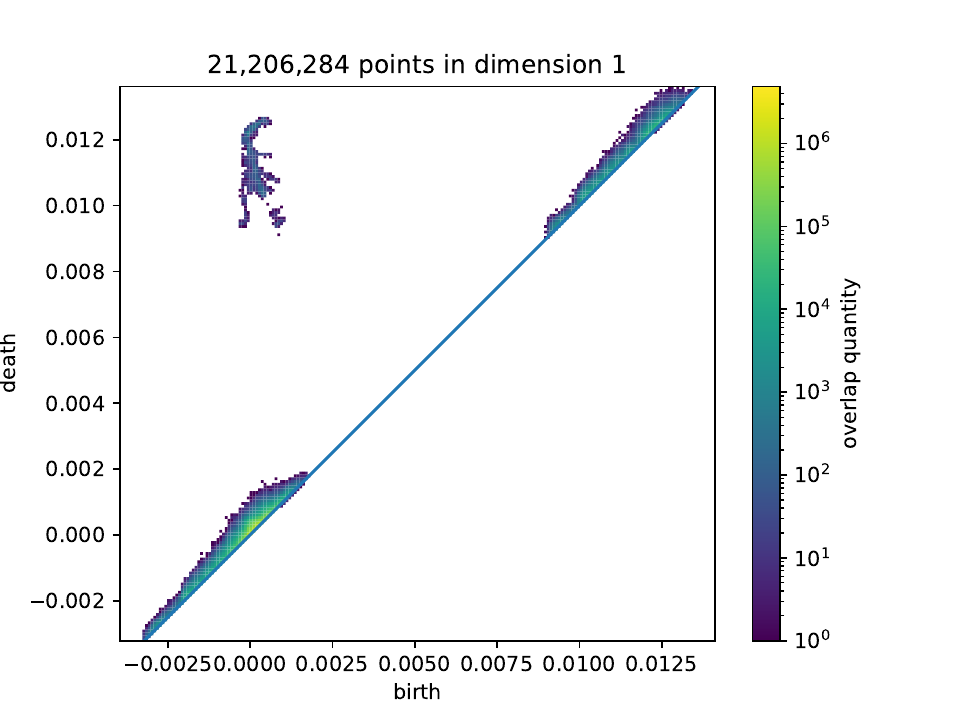}
    \caption{Persistence diagram of the synthetic truss data set in dimension 1.}
    \label{fig:truss_dgm}
\end{figure}

However, the performance is highly dependent on the input, as
\cref{fig:truss_600} shows. The Synthetic Truss dataset 
has very regular and rich topology, as visualized in \cref{fig:truss_vis}.
This figure shows just one particular sublevel set (more precicesely,
its isosurface).
All the $1$-cycles that are shown in this figure will be filled in (die)
at approximately the same threshold value. Moreover, they are all born
at approximately the same threshold value. Hence, whether we distribute
the columns according to the $\low$ value (birth) or to the column value (death), 
we will have a great imbalance in the final number of columns among processors.
Almost all the work will be done by one rank.
The diagram illustrating the distribution of columns at the end of the reduction
is in \cref{fig:truss_col_stats}.
In contrast, \cref{fig:rotstrat_col_stats} shows
that for an example that scales well, the columns are distributed uniformly
on all processors at the end of the reduction.
Persistence diagrams of both data sets are shown in \cref{fig:rotstrat_dgm,fig:truss_dgm}.
Note that the Synthetic Truss diagram has 20 times more points
and they are highly concentrated in one small domain inside the lower left region,
with density higher than $10^6$.

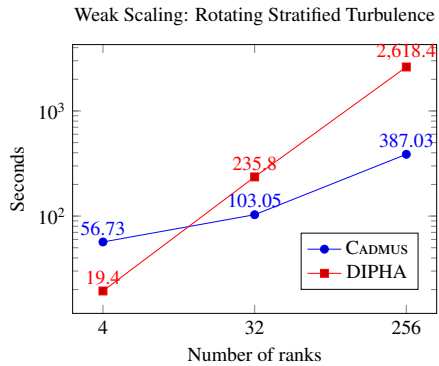
\begin{figure}[]
\centering
\begin{minipage}{0.45\textwidth}
\centering
    \begin{tikzpicture}[scale=0.75]
\begin{axis}[
                title={Weak Scaling: Rotating Stratified Turbulence},
                ymode=log,
                symbolic x coords={4,32,256},
                xtick={4,32,256},
                xticklabels={4,32,256},
                height=2.5in,
                width=.99\textwidth,
                xlabel={Number of ranks},
                ylabel={Seconds},
                nodes near coords,
                point meta=rawy,
                nodes near coords/.style={color=black},
                legend style={at={(0.95,0.3)}},
                legend columns=1,
            ]
    \addplot    +[] table[x=ranks, y=cadmus, col sep = tab]    {data/weak_rotstrat.csv}; \addlegendentry{\cadmus}
    \addplot    +[] table[x=ranks, y=dipha_coh, col sep = tab] {data/weak_rotstrat.csv}; \addlegendentry{\dipha}
\end{axis}
\end{tikzpicture}
    \caption{Weak scaling on the temperature field of rotating stratified turbulence simulation: data sets of size $256^3$, $512^3$ and $1024^3$.}
\label{fig:weak_rotstrat}
\end{minipage}
\end{figure}

It is difficult to evaluate weak scaling for persistence computation for lack of
suitable data. If one takes a large data set and cuts out a small subset, one
inevitably loses large-scale features that often take the most of time to
process. If one coarsens the data set by smoothing, the numerous fine-scale
features get eliminated. Nonetheless, we
investigated the weak scaling properties of our algorithm following the second
procedure. Specifically, we downsampled a $1024^3$ dataset to sizes $512^3$ and $256^3$
by averaging over $2\times 2 \times 2$ voxels.
Dividing the number of processors by $8$ each time, each rank
gets the same amount of input data. However, since persistence diagram also captures global characteristics of the input,
and we effectively smoothed it, removing lots of topological features, one cannot expect
ideal weak scaling. That is indeed the case for the temperature field of turbulence simulation,
plotted in \cref{fig:weak_rotstrat}. The running time is multiplied first by a
factor of $2$; second, by almost
a factor of $4$ for \cadmus.


\section{Conclusion}

We presented a new distributed algorithm for computing persistent
cohomology. It combines space and range partitioning and avoids
computing the Mayer--Vietoris blowup complex. Reducing data locally before
switching to the global reduction is beneficial for scaling, but as always with
topological computation, the performance is data-dependent. For grossly
unbalanced inputs, the computation scales poorly with increasing number of
processors.

Overcoming the data-dependent imbalance of our algorithm is one of the most
pressing issues for future work.  We may also be able to improve the efficiency
of our implementation by experimenting with different data structures for
column addition.  This choice can have a significant impact on the running
time~\cite{phat}.

Although we don't discuss it in this work, persistent cohomology is known to be
particularly efficient for computing persistence of Vietoris--Rips
filtrations~\cite{dualities,ripser}, which makes our algorithm particularly
promising for tackling large problems in that domain. Before doing so, we have
to solve other subproblems involved in that computation, specifically, the
efficient construction of (spacially localized) cliques in the near-neighbor
graphs. We leave this as a direction for future research.

There are applications of persistence that rely on access to
cocycles~\cite{circular-coordinates}. They immediately benefit from our new
algorithm.  At the same time, there are applications that need access to
both cycles and cocycles (as well as their bounding chains and
cochains)~\cite{big-steps}.
In other words, they need both homology and
cohomology computation. Adapting our algorithm to the homology setting is
another important future research direction.
We use the cohomology implementation in \dipha to get a fair comparison, but
experiments with its homology implementation suggest that it is better suited for
the data sets used in this paper.

\section*{Acknowledgment}
This work was supported by the U.S. Department of Energy, Office of Science, Office of Advanced Scientific Computing Research, 
Scientific Discovery through Advanced Computing (SciDAC) program and Mathematical Multifaceted Integrated Capability Centers (MMICCs) 
program, under Contract No. DE-AC02-05CH11231 at Lawrence Berkeley National Laboratory.
This research used resources of the National Energy Research Scientific
Computing Center (NERSC), a U.S. Department of Energy Office of Science User
Facility using NERSC award ASCR-ERCAP-28617.
This research used the Lawrencium computational cluster resource provided by the IT Division at the Lawrence Berkeley National Laboratory 
(Supported by the Director, Office of Science, Office of Basic Energy Sciences, of the U.S. Department of Energy under Contract No. DE-AC02-05CH11231).



\bibliography{references}

\end{document}